\documentclass[superscriptaddress,aps,preprintnumbers,amsmath,showpacs,amssymb,prl,nofootinbib,reprint]{revtex4-1}
\usepackage{bm, color}
\usepackage{amssymb,amsfonts,slashed,amsthm,amsmath,graphicx, soul}
\usepackage{subfigure}
\usepackage{hyperref}
\usepackage{tikz,xcolor}
\usepackage{mathrsfs}

\usepackage{multirow}

\begin{document}

\title{Primordial black hole formation from the merger of oscillons}

\author{Kentaro Kasai}
\affiliation{Institute for Cosmic Ray Research, University of Tokyo, \\
Kashiwa, 277-8582, Japan}

\author{Naoya Kitajima}
\affiliation{Frontier Research Institute for Interdisciplinary Sciences, Tohoku University, \\
6-3 Azaaoba, Aramaki, Aoba-ku, Sendai 980-8578, Japan}
\affiliation{Department of Physics, Tohoku University, \\
6-3 Azaaoba, Aramaki, Aoba-ku, Sendai 980-8578, Japan}


\begin{abstract}
We show that the merger of oscillons results in a broad spectrum of the oscillon mass.
A huge number of oscillon samples obtained from numerical lattice simulations reveal that the oscillon mass distribution has an exponential tail in a heavy-mass region.
This enables us to infer the fractional abundance of heavy oscillons. Using the criterion for the primordial black hole (PBH) formation from the oscillon collapse obtained in previous studies, we estimate the abundance of PBHs and conclude that a sizable number of PBHs can be produced from oscillons. It can be an alternative PBH formation mechanism without employing the tuning of the inflaton potential to enhance the small-scale density fluctuations in the conventional PBH formation scenario.
\end{abstract}

\maketitle


{\bf Introduction.--}
The existence of cosmic inflation in the early universe is now strongly supported by the cosmic microwave background (CMB) observations. According to the observed data, a simple quadratic potential for the inflaton, which is a scalar field driving the inflation, is not favored, implying that the inflaton has self-interactions \cite{Planck:2018jri}. Such self-interactions or interactions with other fields, in general, cause the instability of fluctuations through parametric resonance after inflation, which is called {\it preheating} \cite{Kofman:1994rk,Kofman:1997yn} (see also \cite{Amin:2014eta} for a review). The strong parametric resonance results in a highly inhomogeneous universe soon after inflation, in which the fluctuation exceeds the homogeneous inflaton oscillation and the system enters a fully nonlinear regime.

The oscillon is a quasi-stable non-topological soliton of real scalar field \cite{Gleiser:1993pt,Kolb:1993hw,Copeland:1995fq}.
The oscillons can be copiously produced following the preheating phase if the shape of the inflaton potential is shallower than the quadratic one \cite{Amin:2010dc,Amin:2010xe,Amin:2011hj,Hong:2017ooe,Lozanov:2017hjm}. 
The oscillon has in general a finite lifetime but is typically long-lived \cite{Mukaida:2016hwd,Zhang:2020bec,Zhang:2020ntm,Cyncynates:2021rtf}. Thus, it can change the subsequent thermal history of the universe and can even take a role of dark matter \cite{Olle:2019kbo}.
Oscillons can also be produced from sub-dominant spectator fields, such as moduli and axions (ALPs) \cite{Soda:2017dsu,Kitajima:2018zco,Kawasaki:2019czd,Kawasaki:2020jnw,Fukunaga:2020mvq}.
In addition, the formation of oscillons is accompanied by gravitational wave emissions \cite{Zhou:2013tsa,Soda:2017dsu,Kitajima:2018zco,Lozanov:2019ylm,Hiramatsu:2020obh,Kou:2021bij,Lozanov:2022yoy}, which can be tested by gravitational wave observations. 
See \cite{Kawasaki:2020tbo,Amin:2020vja,Imagawa:2021sxt} for other observational implications of the oscillon.

Another intriguing consequence of the inflation scenario is the formation of primordial black holes (PBHs) \cite{Zeldovich,Hawking:1971ei,Carr:1974nx}.
The PBH formation occurs typically due to the primordial density fluctuations generated during inflation. The scale of the fluctuation is expanded beyond the horizon during inflation, and when the fluctuation reenters the horizon in the subsequent universe and its amplitude exceeds some threshold, it can collapse into a PBH. For reviews, see \cite{Sasaki:2018dmp,Carr:2020gox,Escriva:2022duf}. 
The mass of the PBH depends on the size of the horizon at the formation, and the lifetime is determined by the mass.
If the PBH is long-lived, it can account for all or part of the cold dark matter component today.
On the other hand, light PBHs evaporate in the early epoch and can change the thermal history. In particular, the universe can be successfully reheated through the PBH evaporation even if the inflaton does not interact with the standard model particles.

The oscillon is a nearly-spherical dense object and thus possibly collapses into a black hole. 
The PBH formation from oscillons has been studied in the literature \cite{Kou:2019bbc,Nazari:2020fmk,Muia:2019coe} based on full general relativistic simulations. It has been shown that a single oscillon can collapse into a black hole when its mass exceeds some threshold value with some specific potential form\footnote{
The PBH formation can also result from statistical fluctuations of oscillons as discussed in \cite{Cotner:2018vug}.
}. 

In this letter, we study the PBH formation from oscillons in the cosmological context.
We perform numerical lattice simulations for the scalar (inflaton) field to follow copious oscillon productions. 
Then, we pick up individual oscillons and calculate their masses from simulation data. Finally, we obtain the distribution of the oscillon mass from a large number of oscillon samples.
The resultant oscillon mass distribution shows an {\it exponential tail} due to the merger of oscillons. This enables us to estimate the relative fraction of the heavy oscillons that can collapse into PBHs, and then the PBH formation probability is obtained.
We show that a significant number of PBHs can be formed with natural setups, 
which can explain the dark matter abundance or reheat the universe through the Hawking radiation\footnote{
In Ref.~\cite{Aurrekoetxea:2023jwd}, it turns out that the PBH formation from the oscillon collapse is unlikely in the cosmological setup. Our result does not conflict with this previous result because the PBH formation is still a rare event also in our case and does not occur in typical cases. 
}.


{\bf Model and numerical setup.--} 
Our simulation focuses on the early universe just after the cosmic inflation but before the reheating. In this situation, the inflaton is the only matter content. 
Then, let us consider the model with a single real scalar (inflaton) field, $\phi$, described by the following Lagrangian,
\begin{align}
\mathcal{L} = - \frac{1}{2} \partial^\mu \phi \partial_\mu \phi - V(\phi),
\end{align}
with the potential given by
\begin{align} \label{eq:potential}
V(\phi) = \frac{1}{2\alpha} m^2 \Lambda^2 \left[ \bigg( 1+\frac{\phi^2}{\Lambda^2} \bigg)^\alpha -1 \right],
\end{align}
where $m$ is the inflaton mass near the minimum of the potential, $\Lambda$ is an energy scale associated with the initial amplitude of the inflaton, and $\alpha$ is a constant. 
Assuming the Friedmann-Lema\^{i}tre-Robertson-Walker (FLRW) spacetime with zero spatial curvature,
one can derive the following equation of motion for the inflaton field,
\begin{align} \label{eq:eom}
    \ddot{\phi} + 3H\dot{\phi} - \frac{\nabla^2 \phi}{a^2} + \frac{\partial V}{\partial\phi} = 0,
\end{align}
where $a$ is the scale factor, $H$ is the Hubble parameter, and the overdot
represents the derivative with respect to the cosmic time.  
The energy density of $\phi$ determines the expansion rate of the universe through the Friedmann equation, 
\begin{align} \label{eq:Friedmann}
H^2 = \frac{\rho}{3M_{\rm pl}^2},\quad \rho = \frac{1}{2}\dot{\phi}^2 + \frac{1}{2}(\nabla\phi)^2+V(\phi),
\end{align}
where $M_{\rm pl}\equiv 1/\sqrt{8\pi G}$ is the reduced Planck mass.

On the occurrence of strong parametric resonance after inflation, which is the case of our interest, inhomogeneities grow exponentially, and the system is governed by nonlinear dynamics even if the initial configuration is highly homogenized by the inflationary expansion.
Then, numerical lattice simulations are required to solve the field equation (\ref{eq:eom}) together with the Friedmann equation (\ref{eq:Friedmann}) in the fully nonlinear regime. 

In our simulations, the number of grid points is $N^3 = 512^3$, the initial boxsize is $L=64m^{-1}$, and the comoving lattice spacing is $\Delta x = 0.125m^{-1}$. The scale factor is normalized by the initial value, $a(t_i) = 1$.
The spatial derivative is discretized by the central difference with the second-order accuracy and the dynamical quantities are updated using the second-order leap-frog method with the stepsize $\Delta t = 0.2 \Delta x$.
The initial value of $\phi$ is given by $\phi(t_i,\bm{x}) = \phi_i + \delta\phi(\bm{x})$.
We set $\delta\phi$ at each lattice site by the Gaussian random field generated in the Fourier space with the spectrum of the quantum vacuum fluctuation \cite{Figueroa:2020rrl}.
We set $\dot{\phi}_i=0$ at each lattice point.

To realize strong parametric resonances and subsequent oscillon formations in the present setup, one needs $\Lambda \lesssim 0.05 M_{\rm pl}$ \cite{Amin:2011hj}. Then, throughout the paper, we set $\Lambda = 0.05 M_{\rm pl}$ and $\alpha = 0.15$, but vary $\phi_i$. Note that the dynamics does not depend on the value of $m$ because it can be embedded in the normalization of the spacetime coordinates.

\begin{figure}[tp]
\centering
\includegraphics [width = 8.5cm, clip]{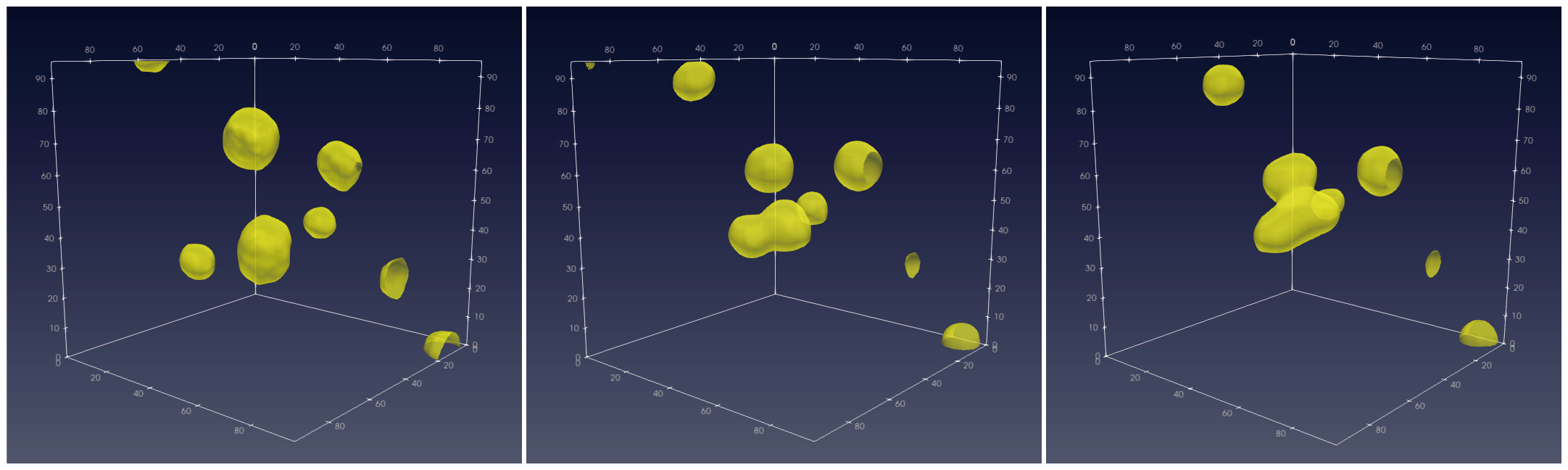}
\caption{
Snapshots of the three oscillon merger process with the time evolving from left to right.
The yellow surface corresponds to the density contrast $\delta = 20$.
The length of each axis is $96\Delta x$.
}
\label{fig:snapshot}
\end{figure}
\begin{figure}[tp]
\centering
\includegraphics [width = 8.5cm, clip]{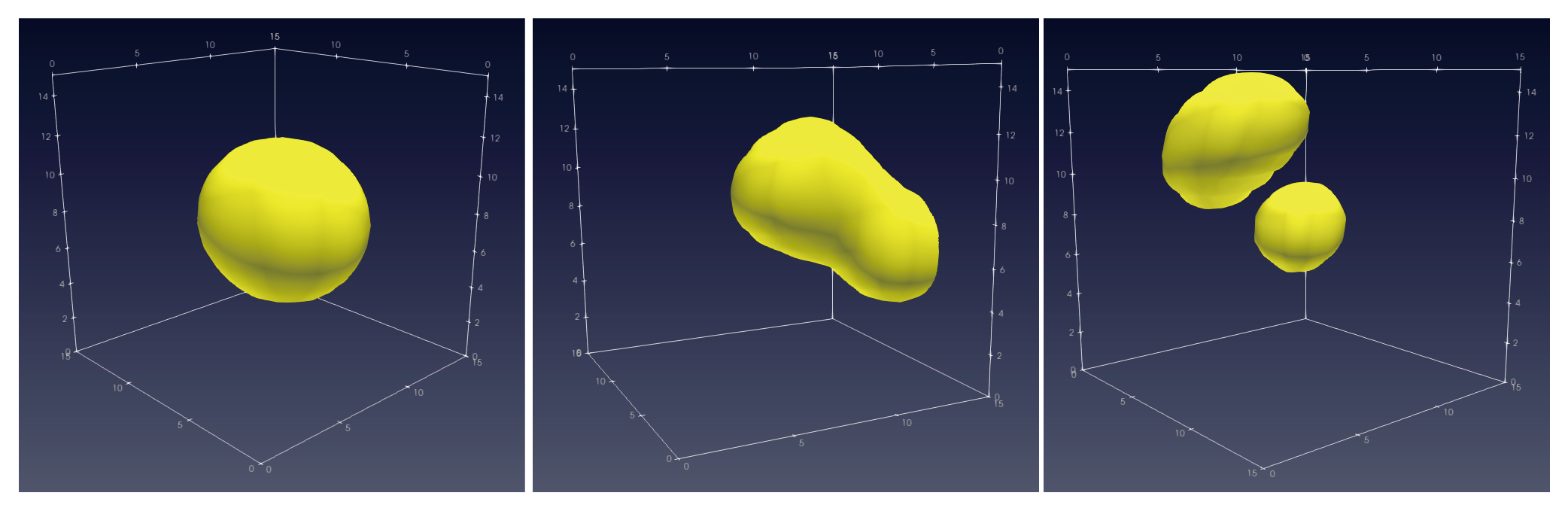}
\caption{
Snapshots of oscillon samples at $t=600m^{-1}$. The yellow surface corresponds to the density contrast $\delta = 500$.
The oscillon masses are $mM/M_{\rm pl}^2 = 2.96$, $2.82$, $2.81$ from left to right.
We have taken $\Lambda = 0.05M_{\rm pl}$, $\alpha = 0.15$ and $\phi_i = 2 \Lambda$.
The length of each axis is $15\Delta x$.
}
\label{fig:oscillons}
\end{figure}

{\bf Oscillon merger.--}
Initial small fluctuations of the inflaton are amplified by parametric resonance instabilities. When the fluctuations become comparable to the background coherent oscillation, filament structures first appear as nonlinear objects. Then, the filament is torn into beads, and they eventually evolve into compact nearly-spherical energy localizations, i.e. oscillons (see e.g. \cite{Kitajima:2018zco}).
Fig.~\ref{fig:snapshot} shows the snapshot of a part of the simulation box after the formation of oscillons. 
Oscillons are actively moving and when two oscillons are closely encountered, they are either merged into a single oscillon or scattered each other, depending on the relative phase difference between two oscillons \cite{Amin:2019ums}.
Fig.~\ref{fig:snapshot} demonstrates the merger of three oscillons near the center of the box.
The merged object, shown in the right panel of Fig. \ref{fig:snapshot}, is highly deformed from the sphere but eventually settles to a nearly spherical object similar to the one in the left panel in Fig.~\ref{fig:oscillons}.

Let us consider the distribution of the oscillon mass as is done in \cite{Kawasaki:2020jnw}.
The mass of each oscillon can be calculated from the simulation by extracting a box enclosing a single oscillon, which we call the {\it oscillon box}.
The first step is to mark the point where the local energy density exceeds the threshold $\rho > 4 \bar\rho$, with $\bar\rho$ being the average energy density in the whole simulation box.
Then, we find the point among those marked points at which the energy density has a peak, $\rho_{\rm peak}$, and identify this point as the center of the oscillon box. The initial size of the oscillon box is $2^3$ in the lattice unit.
The next step is to enlarge the oscillon box by unit lattice size for each side along each axis until the energy densities at every grid points on the surface of the box become lower than the threshold value, $\rho_{\rm th} = 0.05 \rho_{\rm peak}$. 
We discard the oscillon box if it intersects with another one.
Finally, the oscillon mass, $M$, is obtained by taking the sum of the energy density over all points in the oscillon box that satisfy $\rho \geq 0.05 \rho_{\rm peak}$ as follows
\begin{align}
    M = \sum_{\rho > \rho_{\rm th}} \rho a^3 \Delta x^3, \quad \rho_{\rm th} = 0.05 \rho_{\rm peak}.
\end{align}
In order to exclude the large oscillon box that contains multiple oscillons, we set the maximum box size $L_{\rm max} = 20m^{-1}$ for sampling.

We have performed 13,824 simulations with different initial (random) data set, and found total 5,440,258 (7,483,222) oscillon samples for $\phi_i = 1.5\Lambda$ ($2\Lambda$) at the final time of the simulation, $t=600m^{-1}$.
Fig.~\ref{fig:oscillons} shows some samples of the oscillon box. We found that the typical shape of the oscillon is highly spherical as illustrated in the left panel. We also found some samples which are significantly deformed from the sphere as shown in the middle panel. It may be a configuration just after the merger of oscillons. We could not totally exclude sampling the oscillon box that contains multiple oscillons, as shown in the right panel.
Such closely encountered oscillons will be eventually merged into a single oscillon or scattered each other as mentioned above.

\begin{figure}[tp]
\centering
\includegraphics [width = 8.5cm, clip]{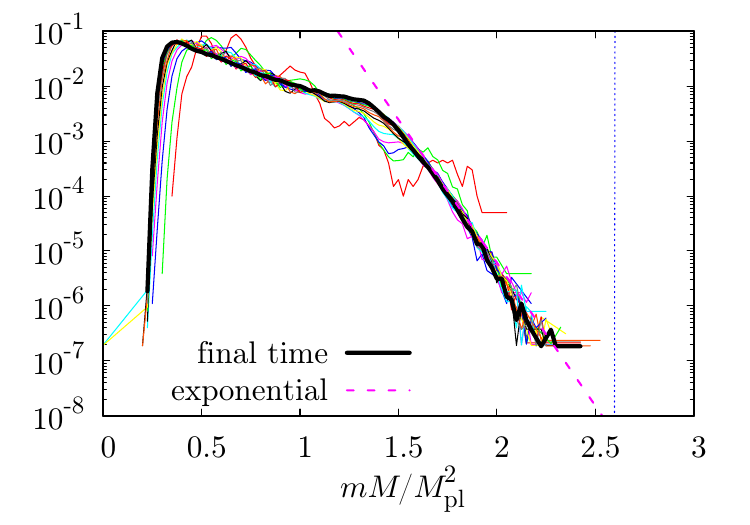}
\includegraphics [width = 8.5cm, clip]{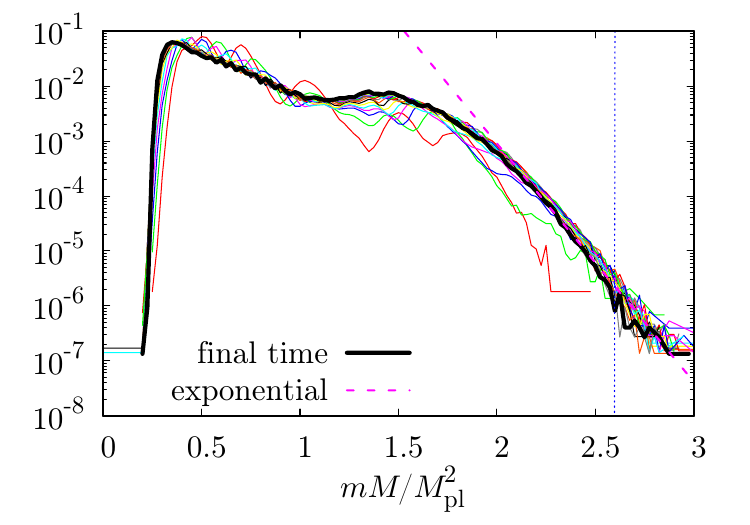}
\caption{
Histogram of the oscillon mass for $\phi_i = 1.5\Lambda$ (top), $2\Lambda$ (bottom). 
The different colors show the different time slices from $t=125m^{-1}$ to $600m^{-1}$ with the interval $\Delta t = 25m^{-1}$. For example, the red, green, blue correspond respectively to $mt=125,\,150,\,175$. The thick black line corresponds to the final time $t=600m^{-1}$.
The dashed magenta line represents the exponential fitting function (\ref{eq:fitting}) 
with $(A,\,B) = (12,\,12)$ (top) and $(13,\,10)$ (bottom).
}
\label{fig:histogram}
\end{figure}

Fig.~\ref{fig:histogram} shows the time evolution of the distribution (histogram) of the oscillon mass, $M$, in units of $M_{\rm pl}^2/m$.
In the early stage, shown in the red, green, and blue curves, the distribution shows an oscillating behavior. It reflects the transient dynamics of the inflaton fragmentation followed by the oscillon formation. After that, the distribution converges to a smooth curve with an exponential tail, denoted by
\begin{align} \label{eq:fitting}
    P(\tilde{M}) = \exp(A - B\tilde{M}),\quad \tilde{M} \equiv \frac{mM}{M_{\rm pl}^2},
\end{align}
where $P(\tilde{M}) {\rm d}\tilde{M}$ denotes the number fraction of oscillons in a range ($\tilde{M},\tilde{M}+{\rm d}\tilde{M}$), and $A$ and $B$ are some numerical constants.
Such an exponential tail might be a result of non-Markovian processes due to the multi-step oscillon mergers.
Note that a similar situation is reported in the ultra-slow roll inflation scenario, where the probability distribution of the curvature perturbation has an exponential tail \cite{Figueroa:2020jkf}.

{\bf PBH formation.--}
The PBH formation from the oscillon collapse with the same potential (\ref{eq:potential}) has been studied in \cite{Kou:2019bbc,Nazari:2020fmk} using full general relativistic simulations.
According to Ref.~\cite{Kou:2019bbc}, the tentative parameter range for the oscillon collapse is $\alpha \sim 0.1$~-~$0.2$ and $\Lambda/M_{\rm pl} \sim 0.05$~-~$0.07$. On the other hand, in Ref.~\cite{Nazari:2020fmk}, viable parameters for 
for the collapse are $\alpha=1/2$, $\Lambda = 0.1 M_{\rm pl}$, $\tilde{M} > 2.63$, or $\alpha=0$, $\Lambda = 0.1 M_{\rm pl}$, $\tilde{M}=2.63,\,2.9$\footnote{
See also \cite{Muia:2019coe} for the case with the $\alpha$-attractor type potential.
In this case, oscillons (oscillatons) collapse into black holes for $\tilde{M} \gtrsim 2.9$.
}.
From the above previous studies, we postulate that the criterion for the PBH formation in our setup ($\Lambda = 0.05M_{\rm pl}$ and $\alpha =0.15$) 
is given by $\tilde{M} >\tilde{M}_{\rm cr}$, with $\tilde{M}_{\rm cr}$ a critical oscillon mass.  
The validity of this assumption should be verified by fully general relativistic simulations, which is left for future work.

The probability for the PBH formation, $\beta$, is defined by the energy fraction of PBHs at the time of formation, 
which can be expressed as
\begin{align} \label{eq:beta}
\beta = 
\frac{\gamma f_{\rm osc}}{\langle \tilde M \rangle}\int^{\infty}_{\tilde{M}_{\rm cr}} \tilde{M}P(\tilde{M}) d\tilde{M},
\end{align}
where $f_{\rm osc}$ is the energy fraction of the oscillon to the total energy which is typically $O(1)$.
The factor $\gamma$ is defined through the PBH mass $M_{\rm PBH} = \gamma M$, quantifying the energy loss of the oscillon at the gravitational collapse and 
$\langle \tilde M \rangle = \int^\infty_0\tilde{M}P(\tilde{M})d\tilde{M}$ is the mean oscillon mass.
Suppose that the probability distribution function in the integral of Eq.~(\ref{eq:beta}) can be approximated by the exponential function (\ref{eq:fitting}), one obtains
\begin{align}
\beta &\approx \frac{\gamma f_{\rm osc}(1+\tilde{M}_{\rm cr}B)}{\langle \tilde{M} \rangle B^2} \exp(A - \tilde{M}_{\rm cr} B).
\end{align}
The values of $\beta$ for various cases are summarized in Table~\ref{tab:beta}.

\begin{table}
\begin{center}
\begin{tabular}{c c c c c}
\hline
~~$\phi_i/M_{\rm pl}$~~ & ~~$(A,B)$~~ & ~~$\langle \tilde{M} \rangle$~~ & ~~ $\tilde{M}_{\rm cr}$~~ & ~~$\beta/(\gamma f_{\rm osc})$~~ \\
\hline
\multirow{2}{*}{$1.5$} & \multirow{2}{*}{$(12,12)$} & \multirow{2}{*}{$0.62$} & $2.6$ & $1.7 \times 10^{-9}$\\
 & & & $3$ & $1.6 \times 10^{-11}$ \\
 \hline
 \multirow{2}{*}{$2$} & \multirow{2}{*}{$(13,10)$} & \multirow{2}{*}{$0.69$} & $2.6$ & $8.8 \times 10^{-7}$ \\
 & & & $3$ & $1.9 \times 10^{-8}$ \\
\hline
\end{tabular}
\end{center}
\caption{The PBH formation probability for various cases. We have taken $\Lambda = 0.05M_{\rm pl}$ and $\alpha = 0.15$.}
\label{tab:beta}
\end{table}


Specifying the oscillon mass, $\tilde{M}$, and the inflaton mass, $m$, the PBH mass is determined as follows
\begin{align}
    M_{\rm PBH} & \simeq 5.3 \times 10^{-12} M_{\odot} \,\gamma \tilde{M} \left(\frac{1\,{\rm eV}}{m} \right) \nonumber \\[1mm]
    & \simeq 1.1\,{\rm kg}\,\gamma \tilde{M} \left(\frac{10^{10}\,{\rm GeV}}{m} \right).
\end{align}
Here, we assume that the inflaton decays into radiation (in the standard model or dark sector) with the decay rate $\Gamma_\phi$. In our setup, the decay rate should be smaller than the inflaton mass,  $\Gamma_\phi < m$, which puts the upper bound on the temperature after the inflaton decay,
\begin{align}
    T_{\rm dec} \lesssim 8.6 \times 10^4\,{\rm GeV}\,g_*(T_{\rm dec})^{-1/4} \left(\frac{m}{1\,{\rm eV}} \right)^{1/2}.
\end{align}
where $g_*$ is the effective relativistic degrees of freedom.


Let us consider the case in which PBHs constitute all dark matter.
Observational constraints indicate that the mass of those PBHs should be within the range of $10^{-16}M_{\odot}\lesssim M_{\rm PBH}\lesssim 10^{-11}M_{\odot}$~\cite{Carr:2020gox} corresponding to the inflaton mass range $1\,{\rm eV} \lesssim m \lesssim 0.1\,{\rm MeV}$.
Using $\rho_{\rm PBH}/\rho_\phi = {\rm const.}$ and $\rho_{\rm PBH}/s = {\rm const.}$ before and after the inflaton decay, with $\rho_\phi$ and $s$ being respectively the energy density of the inflaton and the entropy density, one obtains the current density fraction of the PBH to the total dark matter component,
\begin{align}
    f_{\rm PBH} = \frac{\Omega_{\rm PBH}}{\Omega_{\rm DM}} \simeq 1.7 \times 10^9 \beta \left(\frac{g_*(T_{\rm dec})}{g_{*s}(T_{\rm dec})}\right)\left(\frac{T_{\rm dec}}{1\,{\rm GeV}} \right),
\end{align} 
where $g_{*s}$ is the effective relativistic degrees of freedom for the entropy density, and we have substituted $\Omega_{\rm DM}h^2 \simeq 0.12$ \cite{Tristram:2023haj}.
We can easily realize $f_{\rm PBH} = 1$ with a natural choice of parameters.


Next, we consider an alternative case in which the PBH is evaporated through the Hawking radiation.
In this scenario, we assume that the PBH dominates the universe after the inflaton decay and the reheating results from the PBH evaporation~\cite{Hooper:2019gtx,Inomata:2020lmk}.
The decay rate of the PBH is given by~\cite{Hawking:1978jn}
\begin{align}
    \Gamma_{\rm eva}
    \equiv
    -\frac{{\rm d}\ln M_{\rm PBH}}{{\rm d}t}
    =\frac{\pi\mathcal{G}g_{H*}M_{\rm pl}^4}{480M_{\rm PBH}^3},
\end{align}
where $\mathcal{G}\simeq 3.8$ is the gray-body factor, and $g_{H*}$ is the spin-weighted effective degrees of freedom of thermally produced particles with the Hawking temperature $T_{\rm H}= M_{\rm pl}^2/M_{\rm PBH}$ \cite{Hooper:2019gtx}. 
The condition for the PBH domination at the time of the evaporation (i.e. $H \sim \Gamma_{\rm eva}$) reads
\begin{align}
    \beta &\gtrsim 6.1 \times 10^{-10} \left( \frac{g_{H*}}{g_{*{\rm SM}}} \right)^{1/2}
    \left( \frac{g_{*{\rm SM}}}{g_*(T_{\rm eva})} \right)^{1/4} \nonumber \\[2mm]
    &\quad \times \left( \frac{10^{15}\,{\rm GeV}}{T_{\rm dec}} \right)
    \left( \frac{1\,{\rm kg}}{M_{\rm PBH}} \right)^{3/2},
\end{align}
with $g_{*{\rm SM}} = 106.75$. 
In this scenario, the reheating temperature is given by the PBH evaporation temperature
\begin{align}
    T_{\rm eva} &\simeq 1.1 \times 10^5\,{\rm GeV}\,\left( \frac{g_{H*}}{g_{*{\rm SM}}} \right)^{1/2}
    \left( \frac{g_{*{\rm SM}}}{g_*(T_{\rm eva})} \right)^{1/4} \nonumber \\[2mm]
    &\quad \times \left( \frac{3}{\gamma\tilde{M}} \right)^{3/2} \left( \frac{m}{10^{10}\,{\rm GeV}} \right)^{3/2}.
\end{align}
This PBH reheating scenario 
is viable with a wide range of parameters.


{\bf Discussion.--}
We have shown that oscillons can experience multi-step mergers, producing heavier oscillons. Such heavy oscillon productions result in an exponential tail in the oscillon mass distribution, and it enables us to infer the fractional abundance of oscillons with arbitrarily large masses. Then, based on the previous studies \cite{Kou:2019bbc,Nazari:2020fmk} that demonstrate the PBH formation from oscillon collapses, we have shown that a sizable number of PBHs can be formed from heavy oscillons.
It provides an alternative PBH formation mechanism that does not require the tuning of the inflaton potential to enhance the small-scale fluctuations.

Our analysis relies on the previous studies based on the simulation with the spherical symmetry and the static oscillaton initial condition. 
However, the configuration can be significantly deformed from the sphere, especially after the oscillon merger. In addition, the merger process is highly dynamical and thus the static initial conditions are not suitable.
Therefore, the simulation of the gravitational oscillon collapse should be initiated from more realistic conditions. 
Moreover, the gravitational self-attracting force between oscillons \cite{Ikeda:2017qev,Amin:2019ums} may enhance oscillon mergers or make oscillon binaries. Therefore, it can change the oscillon mass distribution and hence the PBH formation probability.

In addition, highly non-spherical initial configurations may result in nonzero spins for produced PBHs. Thus, it can be a characteristic signature of the PBH formation from oscillons, since the ordinary PBH formation scenario from density fluctuations predicts small spins \cite{Chiba:2017rvs,Harada:2017fjm,Mirbabayi:2019uph,DeLuca:2019buf,Harada:2020pzb,Saito:2023fpt,Saito:2024hlj,Ye:2025wif}.
Calculating the angular momentum of the oscillon is required to predict the spin distribution of PBHs.

The stochastic gravitational wave background can also be a signal of this scenario.
The spectrum may be different from the one induced by the second-order curvature perturbations, which necessarily accompany the PBH formation from primordial curvature perturbations \cite{Saito:2008jc,Saito:2009jt}. Detailed analysis is left for future work.


{\bf Acknowledgments.--}
This work is supported by JSPS KAKENHI Grant Nos. 25KJ1164 (K.K.). 
K.K. was supported by the Spring GX program of the University of Tokyo.
This work used computational resources of Fugaku supercomputer, provided by RIKEN Center for Computational Sciences, through the HPCI System Research Project (Project ID: hp250177).

\bibliography{ref.bib}

\end{document}